\documentstyle[prl,aps,multicol,epsf]{revtex}
\begin{document}
\draft
\widetext
\title{Ferromagnetic transition in a 
double-exchange system containing impurities}
\author{Mark Auslender$^1$ and Eugene Kogan$^2$}
\address{$^1$ Department of Electrical and Computer Engineering,
Ben-Gurion University of the Negev,
P.O.B. 653, Beer-Sheva, 84105 Israel\\
$^2$ Jack and Pearl Resnick Institute 
of Advanced Technology,
Department of Physics, Bar-Ilan University, Ramat-Gan 52900, 
Israel}
\date{\today}
\maketitle
\begin{abstract}
\leftskip 54.8pt
\rightskip 54.8pt
We study ferromagnetic transition in three-dimensional double-exchange model
containing impurities.
The influence of both spin fluctuations and impurity potential 
on conduction electrons is described in coherent potential approximation. 
In the framework of thermodynamic approach we construct Landau functional for 
the system "electrons (in disordered environment) + core spins". 
Analyzing the Landau functional we calculate the temperature of ferromagnetic 
transition $T_C$ and paramagnetic susceptibility $\chi$. For $T_C$, we thus
extend the result obtained by Furukawa in the framework of the Dynamical Mean 
Field Approximation, with which our result coincides in the limit of zero 
impurity potential. We find, that the alloy disorder, able to produce a gap 
in density of electron states, can substantially decrease $T_C$ with respect 
to the case of no impurities.
\end{abstract}
\begin{multicols}{2}
\narrowtext

\section{Introduction}

The recent rediscovery of colossal magnetoresistance in doped Mn oxides
with perovskite structure R$_{1-x}$D$_x$MnO$_3$ (R is a rare-earth metal and D
is a divalent metal, typically Ba, Sr or Ca) \cite{helmolt} 
substantially increased  interest  in the
double-exchange (DE) model \cite{zener,anderson}. Several approaches were 
used lately to study the thermodynamic properties of the DE model, including 
the Dynamical Mean Field Approximation (DMFA) \cite{furukawa2} (and references therein; 
DMFA itself see \cite{DMFA}), Green functions decoupling techniques
\cite{green}, Schwinger bosons \cite{arovas}, 
variational mean-field approach \cite{alonso} and numerical methods 
\cite{yunoki,motome,roder}. In all these approaches chemical
disorder introduced by doping, which is generic for the manganites, has not been 
taken into account. 

Recently we have shown that the
concurrent action of the chemical and magnetic disorder,  is crucial for the
the description of the density of states and conductivity in manganites
\cite{auskog}. 
In the present paper we consider the ferromagnetic transition in the case of
non-zero potential of randomly distributed impurities, using the same 
coherent potential approximation (CPA) \cite{ziman,soven,kubo} as in our
previous paper \cite{auskog} (see also relevant Ref.\cite{letfulov}).
Briefly, the effect of impurities on the transition is the reduction of $T_C$ 
as the impurity potential strength increases. In more detail, at values of the 
impurity potential at which 
the electron band is split off into two sub-bands, the mechanism of ferromagnetic 
exchange is other than in the case of the zero or weak potential.

\section{Hamiltonian and CPA equations}

We consider the DE model with the inclusion of the 
single-site impurity potential. We apply the quasiclassical 
adiabatic approximation and consider each 
core spin as a static vector
of fixed length $S$ (${\bf S}_i=S{\bf n}_i$, where ${\bf n}_i$
is a  unit vector).
The Hamiltonian  of the model $H([{\bf n}_i])$ in site representation is
\begin{eqnarray}
\label{ham}
\hat{H}_{ij}=t_{i-j}+ 
\left(\epsilon_i -J{\bf n}_i\cdot\hat{\bf \sigma}\right)\delta_{ij},
\end{eqnarray}
where $t_{i-j}$ is the electron hopping, $\epsilon_i$ is the 
on-site energy, $J$ is the effective 
exchange 
coupling between a core spin and a conduction electron and 
$\hat{\bf \sigma}$ is the vector of the Pauli matrices. The hat above the
operator reminds that  it is a 
$2\times 2$ matrix in the spin space (we discard the
hat when the operator is a scalar matrix in the spin space).  
The Hamiltonian (\ref{ham}) is random due to randomness of 
a core spin configuration $[{\bf n}_j]$ and the randomness of the on-site energies 
$\epsilon_i$. 

To handle CPA we present Hamiltonian (\ref{ham})  as
\begin{eqnarray}
\hat{H}=\overbrace{\left(t_{i-j}+\hat{\Sigma}\delta_{ij}\right)}^{\hat{H}^0}+
\overbrace{\left(\epsilon_i -J{\bf n}_i\cdot\hat{\bf \sigma}-\hat{\Sigma}
\right)\delta_{ij}}^{\hat{V}}
\end{eqnarray}
(the site independent self-energy $\hat{\Sigma}(E)$ is to be determined later),
and construct a perturbation theory with respect to random potential 
$\hat{V}$.
To do this let us introduce the exact $T$-matrix as the solution of the equation
\begin{equation}
\hat{T}=\hat{V}+\hat{V}\hat{G}_0\hat{T},
\end{equation}
in which $\hat{G}_0=(E-\hat{H}^0)^{-1}$.
For the exact Green function $\hat{G}$ we get
\begin{equation}
\label{green}
\hat{G}=\hat{G}_0+\hat{G}_0\hat{T}\hat{G}_0.
\end{equation}
The self energy is determined from the requirement 
\begin{equation}
\label{gg}
\left\langle\hat{G}\right\rangle_{{\bf n},\epsilon}=\hat{G}_0
\end{equation}
where the angular brackets with the indexes mean averaging over the
configurations of both core spins and impurities.
In CPA $\hat{T}$  is considered in a single-site approximation, at which 
Eq. (\ref{gg}) is reduced to the equation 
\begin{equation}
\label{g}
\left\langle\hat{T}_i\right\rangle_{{\bf n},\epsilon}=0,
\end{equation}
where $\hat{T}_i$ is the solution of the equation
\begin{equation}
\hat{T}_i=\hat{V}_i+\hat{V}_i
\hat{g}(E)\hat{T}_i.
\end{equation}
Here
\begin{eqnarray}
\label{locator}
\hat{g}(E)=g_{0}(E-\hat{\Sigma}), \\
g_{0}(E)=\int_{-\infty}^{\infty}\frac{N_0(\varepsilon)}{E-\varepsilon}d\varepsilon,
\end{eqnarray} 
$N_0(\varepsilon)$ being the bare (i.e. for $\epsilon_i = 0$ and $J=0$) 
density of states  (DOS). Finally Eq. (\ref{g}) can be transformed to an algebraic 
equation for the
$2\times 2$ matrix $\hat{\Sigma}$:
\begin{equation}
\label{gencpa}
\hat{g}=\left\langle\left(\hat{\Sigma}+\hat{g}^{-1}-\hat{V}_i\right)^{-1}
\right\rangle_{{\bf n},\epsilon}.
\end{equation}

\section{Band structure}

To consider the evolution of the DOS 
\begin{equation}
\label{dos}
N(E)=\frac{1}{\pi}\mbox{Im}\;g_c,
\end{equation}
where $g_c=\mbox{Tr}\;\hat{g}$ (here $\mbox{Tr}$ means the trace over spin states only),
with the variation of the impurity concentration and potential strength, we exploit
the semi-circular (SC) bare DOS given at $\left| \varepsilon\right| \leq W$
($W$ is half of the bandwidth) by
\begin{equation}
\label{scdos}
N_{0}( \varepsilon) =\frac{2}{\pi W}
\sqrt{1-\left( \frac{\varepsilon}{W}\right) ^{2}},\;
\end{equation}
and equal to zero otherwise. For this DOS (which is exact on a Caley tree)
\begin{equation}
\label{gf}
g_0(E) =\frac{2}{W}\left[\frac{E}{W}-
\sqrt{\left( \frac{E}{W}\right)^{2}-1}\right].  
\label{scgrfun}
\end{equation}
Hence we obtain
\begin{equation}
\label{sigma}
\hat{\Sigma}=E-2w\hat{g}-\hat{g}^{-1},
\end{equation}
where $w=W^2/8$.
Thus, Eq. (\ref{gencpa}) transforms to 
\begin{equation}
\label{basic}
\hat{g}=\left\langle\left(E-\epsilon+J{\bf n}{\bf \sigma}
-2w{\hat g}\right)^{-1}
\right\rangle_{{\bf n},\epsilon}.
\end{equation}
It is convenient to write the locator $\hat{g}$ in the form
\begin{equation}
\hat{g}=\frac{1}{2}(g_c\hat{I}+{\bf g}_s\hat{{\bf \sigma}}),
\end{equation}
where $\hat{I}$ is a unity matrix. For the charge locator $g_c$ 
and spin locator ${\bf g}_s$ we
obtain the system of equations
\begin{eqnarray}
\label{basic2}
g_c=2\left\langle\frac{E-\epsilon-wg_c}{\left(E-\epsilon-wg_c\right)^2-
(J{\bf n}-w{\bf g}_s)^2}
\right\rangle_{{\bf n},\epsilon}\nonumber\\
{\bf g}_s=2\left\langle\frac{w{\bf g}_s-J{\bf n}}{\left(E-\epsilon-wg_c\right)^2-
(J{\bf n}-w{\bf g}_s)^2}
\right\rangle_{{\bf n},\epsilon}.
\end{eqnarray}

In the strong Hund coupling limit 
($J\rightarrow \infty $) 
we obtain from Eqs. (\ref{basic2}) two decoupled spin sub-bands. For the lower
sub-band, after shifting the energy by $J$ we obtain
\begin{eqnarray}
\label{system}
g_c=\left\langle\frac{1}
{E-\epsilon-wg_c-wn^zg_s}\right\rangle_{{\bf n},\epsilon}\nonumber\\
g_s=\left\langle\frac{n^z}
{E-\epsilon-wg_c-wn^zg_s}\right\rangle_{{\bf n},\epsilon},
\end{eqnarray}
where  ${\bf g}_s=(0,0,g_s)$ an axis OZ is directed along the average magnetization of core 
spins ${\bf m}$).

In fact, details of alloying define how to average over the configurations of impurities 
in Eqs. (\ref{basic2},\ref{system}). We use for this random substitution model of disorder. 
That is $\epsilon_i=V$ with the probability $x$ and $\epsilon_i=0$ with the probability 
$1-x$, where $V$ and $x$ are the impurity potential and concentration, respectively. 
As to core spins, once CPA is introduced, their configuration probability should be 
determined self-consistently in order to close Eqs. (\ref{basic2},\ref{system}). 
We have proved elsewhere that in the presence of an annealed disorder (including 
dynamical one)  DMFA ansatz for the disorder configuration probability 
\cite{DMFA}  keeps the free energy stationary against variations of CPA 
$\hat{\Sigma}$ \cite{auskog1}.

For the following two cases, however, only the averaging over $\epsilon$ is left. 
For a saturated ferromagnetic (FM) phase ($m=1$), we obtain $g_s=g_c$, 
and closed equation for the charge locator is 
\begin{eqnarray}
\label{sat}
g_c=\left\langle\frac{1}
{E-\epsilon-2wg_c}\right\rangle_{\epsilon},
\end{eqnarray}
For a paramagnetic (PM) phase ($m=0$), we obtain ${\bf g}_s=0$, and 
closed equation for the charge locator is 
\begin{equation}
\label{par}
g_c^{(0)}=\left\langle\frac{1}{E-\epsilon-wg_c^{(0)}}\right\rangle_{\epsilon}.
\end{equation}
It appears that calculation of $g_c^{(0)}$ is sufficient for obtaining $T_C$ and 
the PM susceptibility $\chi$ as functions of $x$ and $V$. 

The results of numerical calculation of the DOS at the PM state are 
presented on Fig. 1 
(specific $x$ and different $V$).
\begin{figure}
\epsfxsize=3.5truein
\centerline{\epsffile{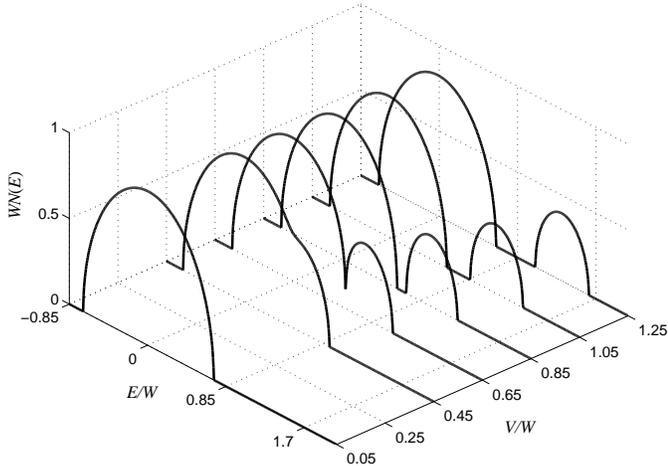}}
\label{Fig.1}
\caption{ The calculated DOS as
function of the relative strength of impurity potential $V/W$ for $x=0.18$.} 
\end{figure}
and on Fig. 2 (specific $V$ and different $x$).
\begin{figure}
\epsfxsize=3.5truein
\centerline{\epsffile{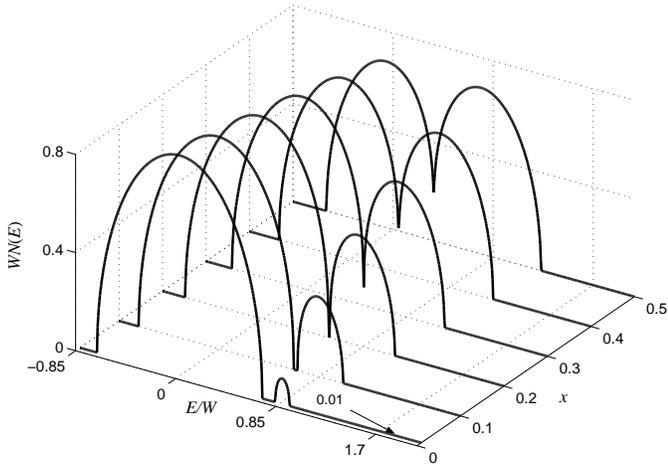}}
\label{Fig.2}
\caption{The calculated DOS as
function of impurity concentration $x$ for $V/W=0.69$.} 
\end{figure}
When comparing Eqs. (\ref{sat}) and (\ref{par}) it is seen that the DOS in
the FM phase is equal to the DOS in the PM state of another model, with increased 
by a factor of $\sqrt{2}$ bare bandwidth. Using this property and Fig. 1 we 
conclude that at an appropriate $V/W$ there may be a gap between  the
conduction band states and the impurity band states in the PM phase, while
the FM DOS is gapless. This may explain  metal-insulator transition  observed in 
manganites and magnetic semiconductors \cite{auskog}.
It is also seen from Fig. 2 that the gap in the DOS existing at low concentrations 
may close at higher concentrations. The value of the impurity potential 
$V_c=W/\sqrt{2}$ detaches two types of the gap behavior. At $V<V_c$ the gap opens 
for some $x<0.5$ or does not open at all, at $V > V_c$ the gap exists for all 
$0<x\leq 0.5$, and at $V=V_c$ the gap closes exactly for $x=0.5$. Anyhow, even if 
the gap is closed there still may exist pseudogap at a strong enough potential 
(see Fig. 2).

The dependence of the gap $\Delta$ on $x$ at several $V/W$ is shown on on Fig. 3, 
which summarizes the regularities discussed above. 
\begin{figure}
\epsfxsize=3truein
\centerline{\epsffile{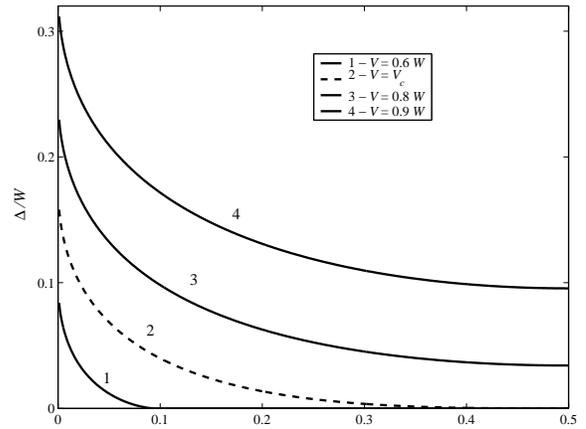}}
\label{Fig.3}
\caption{The calculated gap in the DOS as
function of impurity concentration $x$ at different  strengths of impurity
potential $V$.}
\end{figure}
It is worth noting, that at $V > V_c$, $\Delta$ is near to saturate after 
$x>0.25$.
This may explain the near independence of the resistivity activation energy on 
$x$ in the PM phase observed in single crystalline manganites 
\cite{ramirez} (the fact used by some authors to support the polaron scenarios of 
the transport at $T>T_C$). 
 
\section{Landau functional}

In our approach the calculation of thermodynamic properties is based on the
analysis  of the Landau functional. 
We start from 
the exact partition function of the system electrons + core spins 
\begin{equation}
Z=\int\exp\left[-\beta \Omega([{\bf n}_j])+\beta{\bf H}\cdot\sum_{i=1}^N
{\bf n}_i\right]\prod_{i=1}^Nd{\bf n}_j,
\end{equation}
where $N$ is the number of core spins, ${\bf H}$ is magnetic field,
$\beta=1/T$, and $\Omega([{\bf n}_j])$ is the grand canonical potential
of the electron subsystem  
for a given core spin configuration $[{\bf n}_j]$. Our aim is to obtain Landau
functional of the system.

Since we use CPA for electronic properties it is consistent 
to construct the Landau 
functional using a mean-field approach. Mean-field means that we do not 
take into account large-scale fluctuations of the macroscopic magnetization  
${\bf M}=\frac{1}{N}\sum_j{\bf n}_j$, but within CPA we certainly take into account 
microscopic fluctuations of ${\bf n}_i$. In the mean-field approximation all energy 
levels depend only upon ${\bf M}$. Hence we may approximately put
\begin{equation}
\Omega([{\bf n}_j])\approx \Omega(M),
\end{equation}
where $\Omega(m)$ is the grand canonical potential calculated within CPA at a 
non-zero ${\bf m}$. 
Then the partition function can be written as
\begin{equation}
\label{part}
Z=\int \exp\left[-\beta L({\bf M},{\bf H})\right]d{\bf M} ,
\end{equation}
where
\begin{equation}
\label{part2}
L({\bf M},{\bf H})=\Omega(M)-N{\bf H}\cdot{\bf M}-TS(M);
\end{equation}
the quantity
\begin{equation}
\label{eent}
S(M)=\ln\int 
\delta\left({\bf M}-\frac{1}{N}\sum_{i=1}^N{\bf n}_i\right)\prod_{i=1}^Nd{\bf
n}_j
\end{equation}
is the entropy of the core spin subsystem.
So we may identify the functional $L({\bf M},{\bf H})$ in the exponent of 
Eq. (\ref{part}) with the
Landau functional of the whole system \cite{negele}. 

At high temperatures the minimum of $L({\bf M},0)$ is at $M=0$. At low 
temperatures the point $M=0$ corresponds to the maximum
of this functional. Let us expand $L({\bf M},0)$ with respect to $M$
\begin{eqnarray}
L({\bf M},0)=L_0(T)-L_2(T)M^2+\mbox{O}\;(M^4),\nonumber\\
L_2(T)=\Omega_2(T)-TS_2.
\end{eqnarray}
Here the coefficients are defined from the the second-order expansions 
of $S(M)$ 
\begin{equation}
S(M)=S_0-S_2M^2+\dots,
\end{equation}
where $S_2=\frac{2}{3}N$ \cite{negele},
and of $\Omega(m)$
\begin{equation}
\Omega(m)=\Omega_0(T)-\Omega_2(T)m^2+\dots
\end{equation}
which is to be constructed.
Thus the critical temperature $T_C$, below which the functional has 
the minimum at some $M \neq 0$, is defined from the equation 
\begin{equation}
\label{tc}
L_2(T_C)=\Omega_2(T_C)-T_CS_2=0.
\end{equation}

The magnetic susceptibility in the PM phase is given by the equation
\begin{equation}
\label{chi}
\frac{2}{N}\chi(T)=\frac{1}{TS_2-\Omega_2(T)}.
\end{equation}

The grand canonical potential of the electron subsystem is given by
\begin{equation}
\label{f2}
\Omega(m)=-\frac{N}{\pi\beta}\int_{-\infty}^{+\infty}\ln
\left[1+e^{-\beta(E-\mu)}\right]N(E)dE,
\end{equation}
where $N(E)$ is the DOS given by Eq. (\ref{dos}) and $\mu$ is 
the chemical potential determined from the equation
\begin{equation}
\label{fermi}
n=\int_{-\infty}^{+\infty}f(E,\mu)N(E)dE,
\end{equation}
in which $f(E,\mu)$ is the Fermi function and $n=1-x$ is the number of electrons 
per site.

\section{Calculation of $T_c$ and $\chi(T)$}

Due to the constancy of $n$ (or $x$) the calculation 
of $\Omega_2(T)$ requires only the second-order expansion of $N(E)$ with respect 
to $m$, which is obtained via Eq. (\ref{dos}) thus expanding Eq. (\ref{system}) 
\begin{equation}
\label{exp}
g_c=g_c^{(0)}+g_c^{(2)}m^2+\dots,
\end{equation}
Substituting the related result for $N^{(2)}(E)$ into Eqs. (\ref{f2}),  
(\ref{tc}) and (\ref{chi}) , we obtain 
the following equations
\begin{equation}
\label{ftc}
T_C=\Theta(T_C,\mu(T_C)),
\end{equation}
\begin{equation}
\label{fchi}
\frac{3}{N}\chi(T)=\frac{1}{T-\Theta(T,\mu(T))}	
\end{equation}
In these equations $\mu(T)$ is determined from Eq. (\ref{fermi}) 
where $N(E)$ is replaced by its PM value 
$N^{(0)}(E)$, and 
\begin{eqnarray}
\Theta(T,\mu)=\frac{2w}{3\pi\beta}\int_{-\infty}^{\infty}\ln
\left[1+e^{-\beta(E-\mu)}\right]\nonumber\\
\mbox{Im}
\left\{\frac{\left\langle g_{\epsilon}\right\rangle_{\epsilon}\left[\left\langle
g_{\epsilon}^2\right\rangle_{\epsilon}+\frac{w}{3}\left(\left\langle
g_{\epsilon}\right\rangle_{\epsilon}\left\langle
g_{\epsilon}^3\right\rangle_{\epsilon}-\left\langle
g_{\epsilon}^2\right\rangle_{\epsilon}^2\right)\right]}
{(1-w\left\langle
g_{\epsilon}^2\right\rangle_{\epsilon}) \left(1-\frac{w}{3}\left\langle
g_{\epsilon}^2\right\rangle_{\epsilon}\right)^2}\right\}dE,
\end{eqnarray}
where
\begin{equation}
\label{ge}
g_{\epsilon}=\frac{1}{E-\epsilon-wg_c^{(0)}},
\end{equation}
Using Eq. (\ref{par}) and integrating by parts, we obtain
\begin{eqnarray}
\label{theta}
\Theta(T,\mu)=\int_{-\infty}^{\infty}f(E,\mu)\theta(E)dE,	
\end{eqnarray}
where
\begin{eqnarray}
\theta(E)=-\frac{w}{3\pi}
\mbox{Im}\left\{\frac{\left\langle g_{\epsilon}\right\rangle_{\epsilon}^2}
{1-\frac{w}{3}\left\langle g_{\epsilon}^2\right\rangle_{\epsilon}}\right\},	
\end{eqnarray}
The function $\theta(E)$ being integrated with respect to energy 
gives exactly zero. This leads to $T_C(x)=T_C(1-x)$, 
which reflects the particle-hole symmetry of our model irrespective of the
disorder strength. 

In the case where the DOS is smooth, the inequality $W\gg
T_C$ allows us to consider electrons as nearly degenerate. 
In this case both $\mu$ and 
$\Theta(T,\mu)$ do not depend upon $T$, and Eq. (\ref{ftc}) is just a ready
formula for $T_C$. The same is true if there is developed gap. In this
case $\mu$ is near the middle of the gap, so we can substitute $f(E,\mu)$ 
by one, provided that the integration in Eq. (\ref{theta}) extends 
only over the filled sub-band. In both these cases $\chi(T)$ obeys the Curie-Weiss 
law (see Eq. (\ref{fchi})). In the first case the ferromagnetic order is 
mediated mostly by mobile holes that is specific for DE. In the second case the 
concentration of the mobile carriers (both holes and electrons) is exponentially 
small. So effective exchange between core spins is mostly due to virtual 
transitions of electrons from the lower filled to the upper empty sub-band via the 
gap. This mechanism is an analog of super-exchange (SE) acting in the 
system with electron disorder. 

If $\Delta \sim T$ or there is a pseudogap with a strong dip in the DOS, the 
integration in Eqs. (\ref{fermi}) and (\ref{theta}) should take into account 
the tails of $f(E,\mu)$. The exchange in such cases is intermediate between DE and 
SE types.

For the case of no on-site disorder our Eq. (\ref{ftc}) for $T_C$ coincides with 
Eq. (49) of Ref. \cite{furukawa}, obtained in the framework of DMFA and
calculated numerically versus $x$. In this case
we even managed to calculate the integral in Eq. (\ref{theta}) analytically, 
to get for $T_C$:
\begin{eqnarray}
\label{tcw}
T_C =\frac{W\sqrt{2}}{4\pi}\left[ \sqrt{1-y^{2}}-\frac{1}{\sqrt{3}}\tan^{-1} 
\sqrt{3(1-y^{2})}\right],\\
y=\sqrt{2}\mu/W,
\end{eqnarray}
while Eq. (\ref{fermi}) takes the form
\begin{equation}
\label{c}
x=\frac{1}{2}-\frac{1}{\pi}\left(\sin^{-1}y+y\sqrt{1-y^2}\right).
\end{equation}

When the disorder is taken into account the integration in 
Eq. (\ref{theta}-(\ref{ftc}) is done numerically. The results for $T_C(x)$ at different $V/W$ are 
presented on Fig. 4. The upper 
(dashed) curve calculated at $V=0$ is the same as plotted using Eqs. (\ref{tcw}) and 
(\ref{c})
\begin{figure}
\epsfxsize=3.5truein
\centerline{\epsffile{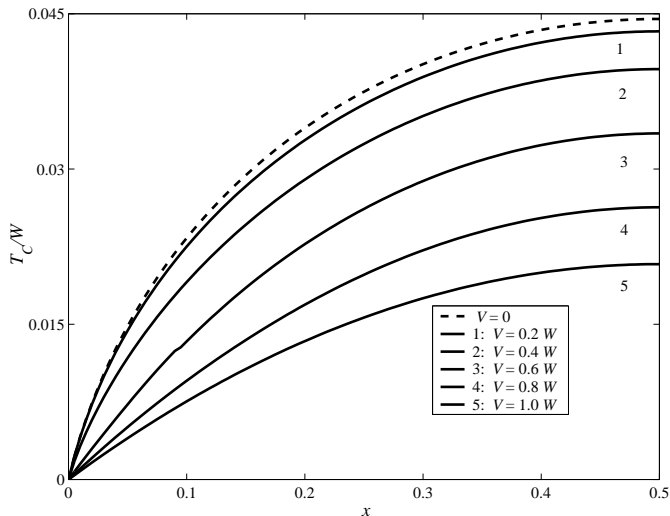}}
\label{Fig.4}
\caption{The calculated Curie temperature as
function of $x$ at different $V/W$.}
\end{figure}

One notices that the increase of the impurity potential leads to progressive decrease of 
$T_C$. This decrease becomes more substantial at $V/W$ and $x$ able to produce the gap 
in the density of states. This trend marks the modification of the 
ferromagnetic exchange mechanism which accompanies the electron band splitting. 
As in the case of DE alone $T_C$ still increases from 
zero to a maximum value upon increasing $x$ from $x=0$ to $x=0.5$.

\section{Discussion}

The  DE scenario of the FM transition in manganites 
attracted especial attention
after its validity was questioned in Ref. \cite{millis}. In the following
discussion, the position of one side can be expressed by reformulation of
the rhetorical 
question by Hubbard \cite{hubbard}:
"How, in the itinerant model, explain a Curie temperature $\sim 1000^\circ$ for
iron, when calculations always give an exchange field $\sim 1-2$ eV?" in the 
form:
How, in the DE model, explain a Curie temperature $\sim 300^\circ$ for
a manganite (at optimal doping), when calculations always give a bandwidth 
$\sim 1-2$ eV? The fallacy is based on the implicit assumption that the $T_C$
should be of the order of the energy difference between the  fully ordered FM
state and the PM state (which is the smallest of the band width
and the exchange integral). It does not take into account that
the fluctuations which
restore magnetic symmetry in itinerant models are local fluctuations 
in the orientation of magnetic
moments, with much lower  energy \cite{hubbard}.
That is why the numerical coefficient in Eq. (\ref{tcw}) turns out to be much
less than one.

Our results show that $T_C$ is further decreased by the
presence of impurities with strong enough potential. It may be questioned in this 
connection why low value of $T_C$ in manganites would be on account of 
fluctuations beyond DMFA in the pure DE model rather than due to the alloy disorder. 
The present study reveals an interesting, though 
hardly experimentally detectable, feature - the alternation of ferromagnetic exchange 
mechanism from DE to SE like as the impurity band splits off the conduction band.   

\section{Acknowledgment}

This research was supported by the Israeli Science Foundation administered
by the Israel Academy of Sciences and Humanities.

\end{multicols}
\end{document}